\documentstyle[prl,aps,multicol,epsf]{revtex}

\newcommand{\unit}[1]{\;\mathrm{#1}}

\begin{document}
\draft

\title{Equilibrium and non-equilibrium electron tunneling via discrete
quantum states}
\author{Mandar M.~Deshmukh, Edgar Bonet, A.~N.~Pasupathy and
D.~C.~Ralph}
\address{Laboratory of Atomic and Solid State Physics, Cornell
University, Ithaca, NY 14853}
\date{\today}
\maketitle

\begin{abstract}
Tunneling is measured via the quantum levels of a metal
nanoparticle. We 
analyze quantitatively the resonance energies, widths, and amplitudes, both in the regime where only one state is accessible for tunneling and in the 
non-equilibrium regime when additional states are made accessible one-by-one.
For tunneling
through one state, our results agree with expectations for 
sequential tunneling, but in the
non-equilibrium regime the resonances are broadened and 
shifted in ways that
require taking into account electron interactions and relaxation.
\end{abstract}

\par

\begin{multicols} {2}
\narrowtext

	In nanometer scale devices, electron transport can occur 
through well-resolved
quantum states.  Transport in semiconductor quantum dots, metal 
nanoparticles, and
molecules can all be understood within a similar framework, in terms 
of the energies of
the states and the rates for transitions between states.  Here we make a quantitative
investigation of 
the fundamental processes at work when electrons tunnel through a 
discrete-state system, using an Al nanoparticle
in a single-electron transistor.  By varying gate 
and source-drain
voltages ($V_g$,$V$), we can manipulate electron flow controllably through one 
state, or through
many, and we can extract tunneling rates for each level.  When only one state participates in tunneling, the
transport properties are in accord with expectations for simple 
sequential tunneling.   However, for larger voltages, the resonance 
energies, widths, and currents can all be modified by the
population of excited quantum states.  
For an
understanding of the high-$V$
regime, non-equilibrium transitions beyond those considered 
previously must be taken
into account.

	A cross-sectional device schematic is shown in 
Fig.~1(a).  The use of
an aluminum particle with aluminum oxide tunnel barriers provides 
mechanical and
charge stability, and allows 
$V$ and $V_g$ to be varied without significantly altering barrier resistances.
Fabrication is done using electron-beam lithography
and reactive ion etching to create a bowl-shaped hole in a silicon-nitride
membrane, with a minimum diameter $\sim10$ nm.  A gate
electrode is formed by depositing 18.5 nm Al, followed by anodization
to 3.5 V in an
oxygen plasma, and then deposition of 8.5 nm of SiO$_x$ 
\cite{dangated}.
The rest of the device is made by depositing a thick Al electrode onto the
bowl-shaped side of the membrane,
oxidizing for 3 min in 50 mTorr of O$_2$, depositing
1.5~nm Al onto the other side of the device to make a layer of 
Al nanoparticles,
oxidizing, and then depositing the lower Al electrode.
Device parameters are determined from the large-$V$ structure of the 
4.2-K Coulomb
staircase curve \cite{chucksthesis}. The capacitance of nanoparticle to the top
electrode is $C_L =7.9 \unit{aF}$, to the bottom electrode
$C_R = 2.7\unit{aF}$,
and the gate capacitance is $C_g=0.06\unit{aF}$.
The sum of the resistances of
\linebreak
\begin{figure}
\vspace{-1.2 cm}
\begin{center}
\leavevmode
\epsfxsize=7 cm
\epsfbox{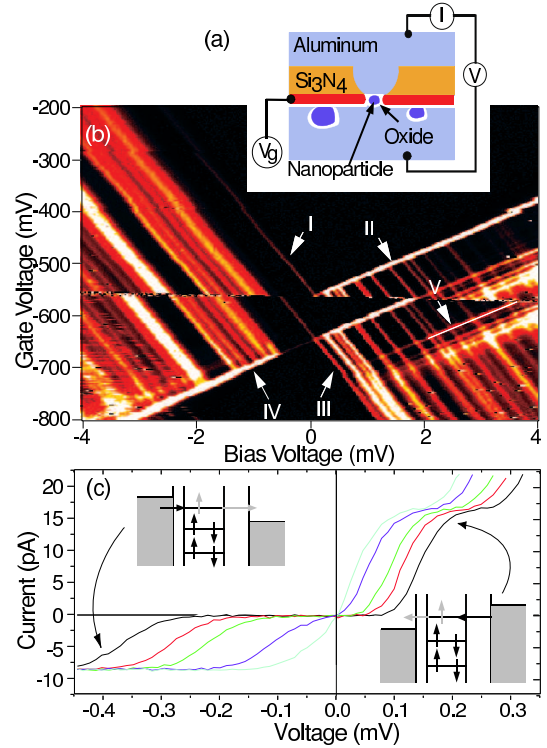}
\end{center}
\caption{
\label{figure1}
(a) Cross-sectional device schematic.
(b) Colourscale differential conductance as a
function of $V_g$ and $V$.  A $0.06$ Tesla field is applied to drive the Al leads normal.
(c) $I$ \textit{vs.}\ $V$ for different $V_g$ in the equilibrium regime. The steps correspond to resonances III and IV. Insets show
the corresponding tunneling transitions. }
\end{figure}
\noindent
the two tunnel junctions is $R_{\Sigma}
\approx 3
M\Omega$,
with individual resistances sufficiently large that the intrinsic
widths of 
the quantum states are smaller than $k_BT$.
Assuming a 
roughly hemispherical \cite{chucksthesis}
particle shape, and a capacitance per unit area of 
$0.05\unit{aF/nm}^2$ \cite{jialu},
we estimate a nanoparticle diameter $\sim 10$ nm.

In Fig.~1(b), we plot the differential conductance 
$dI/dV$ 
as a function of $V_g$ and $V$, when the sample is cooled in
a dilution refrigerator with copper-powder filters on the electrical leads.  The lines in the figure
are due to tunneling
resonances through discrete quantum states in the nanoparticle. 
Lines having a positive
slope correspond to tunneling thresholds across the 
lower-resistance junction L, and negative
slopes are
thresholds across junction R \cite{vdr}.  The 
discontinuity 
evident in the figure is
due to a $V_g$-driven change in the charge on another nanoparticle 
adjacent to the
one through which tunneling occurs.  This merely shifts the
electrostatic potential of the current-carrying particle.  
The intrinsic energies, current
levels, and widths of the resonances are not otherwise altered, so 
that the full $dI/dV$-spectrum can be constructed.
From the absence of spin-Zeeman splitting in a magnetic field (not 
shown) for resonance lines I and II, we can identify 
these transitions with tunneling from an odd number of electrons 
$n_0$ on the particle to an even number \cite{dan}.  Since these  resonances 
require increased $|V|$ as a function of $V_g$, they are 
$n_0\! \rightarrow \! n_0\!-\!1$ transitions.  (See Fig.~2.)
Resonances III and IV correspond to even $(n_0\!-\!1)\! \rightarrow \!$ 
odd $n_0$ transitions.
The large gaps in $V$ between each of resonances I and II and the next
parallel lines are due to the energy difference 
$\sim 2\Delta$
between a fully paired
superconducting state in the Al particle and the next 
lowest-energy tunneling state
with 2 quasiparticles \cite{dangated}.

{\em Equilibrium Regime:}
We first consider the region of $V_g$ near $-650$ mV, 
where $V_g$ and
$V$ can be adjusted so that a single spin-degenerate quantum state (state 0)
is accessible for tunneling ($I$-$V$ curves are shown in 
Fig.~1(c)).  For the case under consideration, in which the 
quantum
level is either empty or singly occupied, the current predicted for sequential
tunneling \cite{sequential,vdr,remark,otherspin} is
\begin{equation}
I = e \frac{2 \gamma_{0L} \gamma_{0R}(f_L - f_R)}{(1+f_L) \gamma_{0L} + (1+f_R)
\gamma_{0R}}
\end{equation}
where $\gamma_{0L}$ ($\gamma_{0R}$) is the bare rate for an electron 
to tunnel from
the quantum state 0 to an unoccupied density of states in electrode L (R)
and $f_i$ is the occupation probability for states in electrode $i$ 
with energy equal to the
resonance state (for a thermal distribution 
$f_i=[1+exp[(\epsilon_0-\mu_i)/kT]]^{-1}$,
with  $\epsilon_0$ the energy to occupy the quantum state, and $\mu_i$ the
chemical potential in electrode $i$).  Our observations
are in excellent accord with this model.  For instance, the tunneling current
through the quantum state is not the same for both bias directions,
being $I_+=16.6 \pm 0.1 {\unit pA}$ for $V\!>\!0$ ({\it i.e.}\ crossing lines II or III)
 and $I_-=-8.4 \pm
0.1
{\unit pA}$ for $V\!<\!0$ (crossing lines I or IV).
This has been
observed previously \cite{cobden,deshpandeprb} and is a consequence of
spin-degeneracy. For $V\!>\!0$, electrons tunnel across the
high-resistance rate-limiting tunnel barrier into an empty state,
so that either spin-up or
spin-down electrons can tunnel.
For $V\!<\!0$, the rate-limiting step is for an
electron of a given spin on the particle to tunnel through
the high-resistance barrier, and the current
level is approximately cut in half.  By equating the measured 
currents to Eq.~(1), we
determine
$\gamma_{0R} = (5.3 \pm 0.1)  \times 10^{7} s^{-1}$ for the
high-resistance
junction 
\linebreak
\begin{figure}
\begin{center}
\leavevmode
\epsfxsize=7 cm
\epsfbox{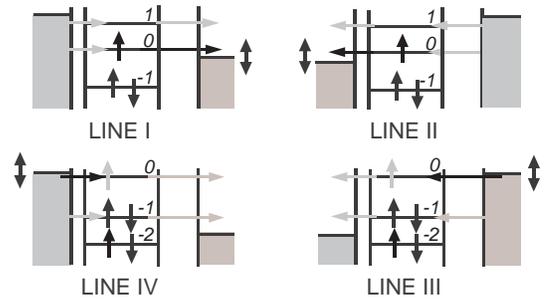}
\end{center}
\vspace{-0.4 cm}
\caption{
\label{figure2}
Tunneling diagrams depicting tunneling transitions active for resonance 
lines I-IV in a non-equilibrium regime.  Black spins represent the 
ground-state electron configuration.  Black arrows indicate the threshold tunneling transition, and gray arrows denote other transitions that contribute to the current for the value of $V$ depicted.}
\end{figure}
\noindent
and $\gamma_{0L} \sim 7 \times 10^{9} s^{-1}$. 
In addition, for resonances ({\it e.g.,} III, IV) in which the spin-degeneracy of the state is split by
a magnetic
field, the currents through the 
two Zeeman states for a given bias 
direction are not equal
\cite{dan,deshpandeprl}, in agreement with the simple tunneling 
theory.  The maximum
current through the lower-energy Zeeman state is $e \gamma_{0L}
\gamma_{0R}/(\gamma_{0L} + \gamma_{0R})$ for either bias direction, and
the second
state then adds current to produce the maximum allowed by Eq.~(1).

An interesting feature of Eq.~(1) when the tunneling threshold is 
across the lower resistance barrier (peaks II and IV) is that the 
maximum of $dI/dV$ does not occur exactly when $\mu_i = \epsilon_0$.  Instead, the resonance is
shifted to lower $|V|$, by an amount $\propto T$, so that it shifts with $T$. This is due 
to charge accumulation on the particle as the Fermi function sweeps by the energy of the 
quantum state, which limits the current at larger $|V|$ on account of Coulomb 
blockade.
Our observations of this effect (not shown) are 
equivalent to
results of Deshpande {\it et al.}\ \cite{deshpandeprb}, although
the data in \cite{deshpandeprb} were compared to an approximation that differed from Eq.~(1) 
\cite{remark}.

{\em Non-equilibrium Regime:}
We can controllably tune the device so that 
more than one quantum
state can participate in tunneling.  This is illustrated by following 
line II in
Fig.~1(b).  This line corresponds to processes which are 
initiated by an electron
tunneling off the nanoparticle from the quantum state 0 to electrode L.
However, as one follows line II to higher V, past negative-sloping 
resonance lines which
intersect line II, these lines indicate that the subsequent tunneling 
of an electron from
electrode R back onto the nanoparticle can proceed via many different 
energy levels other
than state 0.  The total current under these conditions can be 
modeled by a master
equation which takes into account all allowed transitions between the
energetically-accessible $n_0$- and ($n_0\!-\!1$)-electron states \cite{vdr}.
In Fig.~2, we show tunneling diagrams depicting representative accessible states
for resonances I-IV in a non-equilibrium regime.
\linebreak
\begin{figure}
\vspace{-0.75 cm}
\begin{center}
\leavevmode
\epsfxsize=6 cm
\epsfbox{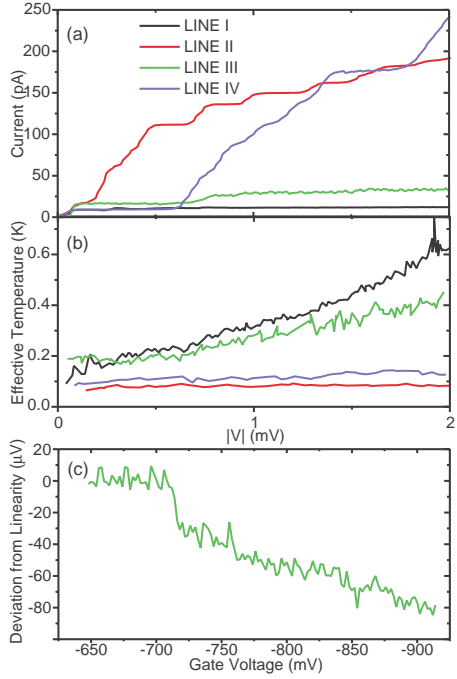}
\end{center}
\caption{
\label{figure3}
Properties of the resonance lines as $V_g$ is varied to change the value of the $V$ at which the resonances appear.  The second tunneling states become energetically accessible for $|V|> 0.18$ mV for resonance line I, $0.19$ for line II, $0.50$ mV for line III, and $0.64$ mV for line IV.
(a) Magnitude of tunneling current. (b) Width of the $dI/dV$ peak as a function of $V$, expressed as an effective temperature.
(c) Deviation from linearity for the $V$-position of peak III as a function of $V_g$.  A line was fit to the peak position in the equilibrium regime between $V_g= -650$~mV and $-701$~mV, and this was subtracted from the measured positions.}
\end{figure}

In Fig.~3(a) we plot the step height in current associated with resonance lines I-IV, as 
$V_g$ and $V$ are tuned to follow the lines in the $|V|$-$V_g$ plane. 
Peaks I and III 
have approximately constant amplitude, while the 
currents for peaks II
and IV grow quickly as $|V|$ enters the non-equilibrium regime. 
This can be understood
trivially.  For peaks I and III, the tunneling threshold is across the 
higher-resistance
tunnel junction R.  This junction is always rate-limiting and it 
matters little how many transport channels
are available across junction L. For
peaks II and IV, the tunneling threshold is across the 
low-resistance barrier L,
but the rate-limiting process is 
how quickly electrons can
tunnel across the other barrier.  As $|V|$ is increased, more quantum 
levels contribute to
this process, and the current grows.  By measuring the 
current along
peaks II and IV as levels are added one by one, and then fitting to master-equation results, we can measure rate-limiting tunneling rates for each
quantum state: the 
$\gamma_{iR}$, for $i=0~\mathrm{to}~5$, are $(5.3, 15.7, 8.0, 16, 15,
9)\times 10^7 s^{-1}$,  $\bar{\gamma_L}\approx 3 \times 10^{9} s^{-1}$,
and (assuming relaxation effects are negligible) $\gamma_{-1R} \approx 17 \times 10^7 s^{-1}$.

Figure 3(b) shows the widths of the tunneling resonance lines I-IV.  These were
determined by fitting the $V$ dependence of each conductance peak to 
the derivative of a
Fermi function, and then converting the  voltage width to effective 
temperature by
multiplying by the capacitance ratio $(e/k_B)C_L/(C_L+C_R)$ for peaks 
I and III or
$(e/k_B)C_R/(C_L+C_R)$ for peaks II and IV \cite{dan}.  In either the
equilibrium or non-equilibrium regimes, the prediction of the 
simplest master equation \cite{vdr}
is that the peak shape
should be a derivative of the Fermi function with a
width approximately equal to the electronic temperature in the electrodes.  Our 
measurements agree with this model within the equilibrium regime, 
with a constant electron temperature $T \approx$ 90 mK.  
This is significantly higher than the $T$ achieved in non-gated tunneling devices using our apparatus, 45 mK, and we ascribe the difference to heating by leakage current from the gate to electrode R. 
Line III is broader than the others at low $V$ because the magnetic field of $60$~mT applied to drive the Al electrodes normal produces an unresolved Zeeman splitting 
($\Delta E/k_B = 2\mu_B H/k_B = 80$~mK).

As $|V|$ is increased into the non-equilibrium regime, 
peaks I and III undergo large increases in width and peak IV broadens slightly,
while peak II shows no measurable change.
The differences are not merely an effect of heating in the electrodes, 
because peaks II and IV have the
largest magnitudes of current and power.
We suggest that these measurements can be
explained as a consequence of electronic interactions in the 
non-equilibrium regime, by a
mechanism due to Agam {\it et al.}\ \cite{agam}.  Consider 
resonance line III, for which the tunneling threshold corresponds to 
an electron entering
quantum state 0.  For $V > 0.50$ mV 
the next tunneling event, which 
discharges the particle, may 
occur out of different,
lower energy states (see Fig.~2), leaving an electron-hole excitation on the 
nanoparticle.  Agam
{\em et al.}\ suggested that if this non-equilibrium state does not 
relax before the next
electron tunnels onto the particle, it can shift the energy of the 
tunneling resonances on
account of an alteration of the electron-electron interaction energy
\cite{agamnote}. 
In past work on
smaller aluminum particles, shifted transitions were resolved 
individually \cite{agam}; however the
relative shift is expected to decrease with increasing nanoparticle 
size \cite{agam}, so it
is reasonable that the shifts would produce only broadened resonances 
for the 10 nm
particle under investigation here.  Because a growing ensemble of 
different non-equilibrium states
can be excited with increasing $|V|$, this mechanism can explain the 
increase in width of line III as a
function of $|V|$.
The same non-equilibrium mechanisms should also come into play for line IV, for $V< -0.64$~mV, but the broadening here is reduced because the threshold tunneling event is across the lower-resistance junction, L.  Barrier R quickly becomes rate limiting as line IV is crossed, so that higher-energy non-equilibrium resonances do not add significant additional current.
In order for the non-equilibrium mechanism to
apply for line III, the relaxation rate of some non-equilibrium excitations to the ground state must be comparable to or slower than
$\gamma_{0R} = 5.3\times 10^{7} s^{-1}$.  The rate predicted
by Agam {\it et 
al.}\ for {\it spin-preserving} energy relaxation in
aluminum particles is $\sim 10^8 s^{-1}$ \cite{agam}.

Resonances I and II are a different case, because the
tunneling threshold corresponds to an electron leaving 
quantum
state 0.  The subsequent 
tunneling event, adding an electron back to the nanoparticle, may for 
large $V$ occur
into higher-energy
states (see Fig.~2(a,b)), but nevertheless this excitation alone cannot produce a
non-equilibrium shift in the energy of subsequent discharging transitions.
The reason is that
only this electron is free to tunnel off the nanoparticle; there is 
no electron in quantum
state 0 whose transition energy might be shifted.
Therefore within the picture of Agam {\it et al.}\ \cite{agam}, no non-equilibrium broadening
should be expected for levels I and II, in conflict with the data for level I.  This discrepancy
can be explained if non-equilibrium excitations on the nanoparticle 
can be generated
not only by the tunneling transitions on or off the particle that have been
considered previously \cite{agam}, but also by transitions in which a 
high-energy
electron relaxes within the nanoparticle and produces an 
electron-hole excitation.
Broadening would then be generated by the Agam
mechanism.  Within this scenario, the difference between the broadening visible for 
resonance line I and the
lack of broadening of line II would follow from the fact that for 
peak II a high-energy
electron on the particle can quickly exit through the low-resistance 
tunnel junction L,
while for peak I the high-energy particle must exit through the 
high-resistance junction R,
giving a much longer residence time during which relaxation 
transitions can occur.
In order for line I to be broadened, the fastest relaxation rates must become comparable
to $\gamma_{0R} = 5.3\times 10^{7} s^{-1}$ as $|V|$
increases.

By tuning $V_g$ and $V$ into
the non-equilibrium regime, the apparent {\em 
energies} of the $dI/dV$
peaks can also be changed.  This is clearest for line III (Fig.~3(c)), which undergoes a shift
of $33 \mu \mathrm{V}$ to lower voltage when the threshold for
non-equilibrium 
tunneling via state -1 (line V in
Fig.~1(b)) is crossed. 
Because we have measured rate-limiting tunneling rates for the energetically-accessible states from the current amplitudes, 
we can test whether this shift can be explained by the simplest master equation \cite{vdr}, which assumes that the underlying energies of the quantum states are not changed by non-equilibrium interactions.  
Only one relevant parameter is not determined previously: $x = \gamma_{0L}/\gamma_{-1L}$.
The solution of the master equation does predict a voltage shift ($\propto\! T$) for the conductance peak compared to the equilibrium case, and for $x< 0.15$ it can explain the full value of the experimental shift.  However, we judge this to be improbable, because the measured values of $\gamma_{iR}$ fall within a more narrow distribution.  For $x\! \sim\! 1$ in Eq.~(2), the predicted shift is $15 \mu \mathrm{V}$ -- much smaller than we measure.
We can more naturally
explain the full value of
the shift by again taking into account that the presence of 
a non-equilibrium
excitation can change the energy of a tunneling 
transition.  In the equilibrium
regime, the occupation of quantum state 0 corresponds to a transition 
from a fully-paired
superconducting state on the aluminum particle to a state with one high-energy
quasiparticle; in the non-equilibrium case, the transition can be 
from a state with two
quasiparticles to one, with a
transition energy lowered by $\sim\!2\Delta\! \approx\!$ 0.35~meV
\cite{janmoshe}.   This is much bigger than $k_BT\! \sim 10\!\mu \mathrm{eV}$, and in this case the observed resonance is shifted to lower $|V|$ by an amount $\propto\! T$ because electrons in the tail of the Fermi distribution can excite the non-equilibrium state and open the lower-energy current channel.
For $x\!=\!1$ and $T\!\sim \!90$~mK the master-equation result is that the measured shift can be
produced by a
non-equilibrium lowering of the
transition energy by any amount greater than $20 \mu \mathrm{eV}\!\sim\! 2k_BT$.

In summary, we have made a systematic study of the transition energies, widths,
and current levels for electrons tunneling through discrete quantum states in a
nanometer-scale electronic device.
When transport occurs through a
single quantum level, the results are in agreement with the
expectations of sequential tunneling.  At large values of $|V|$, the
non-equilibrium population of excited electronic states,
together with electron-electron interactions, can modify the
widths and apparent energies of the tunneling resonances.

Acknowledgements: NSF DMR-0071631, the Packard Foundation,
and the Cornell Nanofabrication Facility.

\end{multicols}

\begin{thebibliography}{99}
\bibitem{otherspin} In the case where one quantum level is either singly or
doubly occupied, the
sequential tunneling current is
\begin{equation}
I = e \frac{2 \gamma_{0L} \gamma_{0R}(f_L - f_R)}{(2-f_L) \gamma_{0L} +
(2-f_R)
\gamma_{0R}}. 
\end{equation}
\bibitem{agamnote} A statement in \cite{agam} that the first tunneling
resonance should be unaffected by non-equilibrium is incorrect in some cases when spin is
taken into account.
\end{thebibliography}
\end{document}